# What is Visualization Really for?

MIN CHEN, University of Oxford
LUCIANO FLORIDI, University of Hertfordshire and University of Oxford
RITA BORGO, Swansea University

Whenever a visualization researcher is asked about the purpose of visualization, the phrase "gaining insight" by and large pops out instinctively. However, it is not absolutely factual that all uses of visualization are for gaining a deep understanding, unless the term insight is broadened to encompass all types of thought. Even when insight is the focus of a visualization task, it is rather difficult to know what insight is gained, how much, or how accurate. In this paper, we propose that "saving time" in accomplishing a user's task is the most fundamental objective. By giving emphasis to saving time, we can establish a concrete metric, alleviate unnecessary contention caused by different interpretations of insight, and stimulate new research efforts in some aspects of visualization, such as empirical studies, design optimisation and theories of visualization.

## 1. INTRODUCTION

*Visualization* was already an overloaded term, long before it has become a fashionable word in this era of data deluge. It may be used in the context of meditation as a means for creative imagination, or in sports as a means for creating a heightened sense of confidence. If we consider the term literally, as Robert Spence said, "visualization is solely a human cognitive activity and has nothing to do with computers" [Spence 2007].

In this article, we focused on visualization in computing, which may be referred to technically as *Computer-supported Data Visualization*. In this context, the process of visualization features both data and computer. These two essential components differentiate this technological topic from those above-mentioned contexts. In the remainder of this article, we will simply refer to "computer-supported data visualization" as "visualization".

Scott Owen [1999] compiled a collection of definitions and rationale for visualization, most of which are still widely adopted or adapted today. These definitions were intended to define the two questions, namely what is visualization and what is it for?

- "The goal of visualization in computing is to gain *insight* by using our visual machinery." [McCormick et al. 1987]
- "Visualization is a method of computing. It transforms the symbolic into the geometric, ... Visualization offers a method for seeing the unseen. It enriches the process of scientific discovery and fosters profound and unexpected *insights*." [McCormick et al. 1987]
- "Visualization is essentially a mapping process from computer representations to perceptual representations, choosing encoding techniques to maximize human understanding and communication." [Owen 1999]
- "Visualization is concerned with exploring data and information in such a way as to gain understanding and *insight* into the data. The goal ... is to promote a

  deeper level of understanding of the data under investigation and to foster new *insight* into the underlying processes, relying on the humans' powerful ability to visualize", [Earnshaw and Wiseman 1992]
- "The primary objective in data visualization is to gain *insight* into an information space by mapping data onto graphical primitives." [Senay and Ignatius 1990]

In addition to Scott Owen's collection, there are other commonly cited definitions:

- Visualization facilitates "the use of computer-supported, interactive, visual representations of abstract data to amplify cognition." [Card et al. 1999]
- "Graphics reveal data. Indeed graphics can be more precise and revealing than conventional statistical computations." [Tufte 2001]
- "Information visualization helps think." [Few 2009]
- "Information visualization utilizes computer graphics and interaction to assist humans in solving problems." [Purchase et al. 2008]
- "The goal of information visualization is to translate abstract information into a visual form that provides new *insight* about that information. Visualization has been shown to be successful at providing *insight* about data for a wide range of tasks." [Hearst 2009]
- "The goal of information visualization is the unveiling of the underlying structure of large or abstract data sets using visual representations that utilize the powerful processing capabilities of the human visual perceptual system." [Berkeley 2010]
- "The purpose of visualization is to get *insight*, by means of interactive graphics, into various aspects related to some processes we are interested in ..." [Telea 2008]

In the above definitions, there are many references to *gaining insight*, or likewise phrases such as *amplifying cognition*, *seeing the unseen*, *unveiling structure*, *answering questions*, *solving problems*, and so forth. It is unquestionable that these are the benefits that visualization can bring about in many occasions. There has been an abundance of evidence to confirm such goals are achievable. However, *insight* is a non-trivial concept. It implies "accurate and deep intuitive understanding" according to many dictionaries. While it is what everyone who creates or uses visualization is inspired to achieve, it is an elusive notion and rather difficult to measure, evaluate, or validate objectively.

Perhaps it is also because of its vagueness, it is relatively easier for people to interpret the term *insight* differently. The charged debate about chart-junks a few years ago was perhaps partly caused by the diverse interpretation of what *insight* to be gained from visualization.

The debate started with a paper by Bateman et al. [2010], which reported an empirical study on the effects of using visual embellishments in visualization. They compared conventional plain charts with highly embellished charts drawn by Holmes [1984]. The findings of the study suggest that embellishment may aid memorization. Following this work, Hullman et al. [2011] proposed a possible explanation that "introducing cognitive difficulties to visualization" "can improve a user's understanding of important information." Obviously this was a major departure from the traditional wisdom of avoiding chart-junks in visualization. For example, in [Tufte 2001], some of Holmes's visual designs were shown as counter examples of this wisdom.

These two pieces of work attracted much discussion in the blogosphere. Stephen Few, the author of several popular books on visualization (e.g., [Few 2009]), wrote two articles. On [Bateman et al. 2010], he concluded:

> "At best we can treat the findings as suggestive of what might be true, but not conclusive." [Few 2011a]

Few was much more critical on [Hullman et al. 2011]:

> "If they're wrong, however, which indeed they are, their claim could do great harm." [Few 2011b]

In many ways, the two sides of the debate were considering different types of insight to be gained in different modes of visualization. We will revisit this debate later in Section 3.2.

**2. A STORY OF LINE GRAPH**

Before we attempt to answer the question what visualization is really for, let us examine some examples of visualization. We start with one of the simplest form of visualization, *line graph*, which is also referred to as *line chart* and *line plot*.

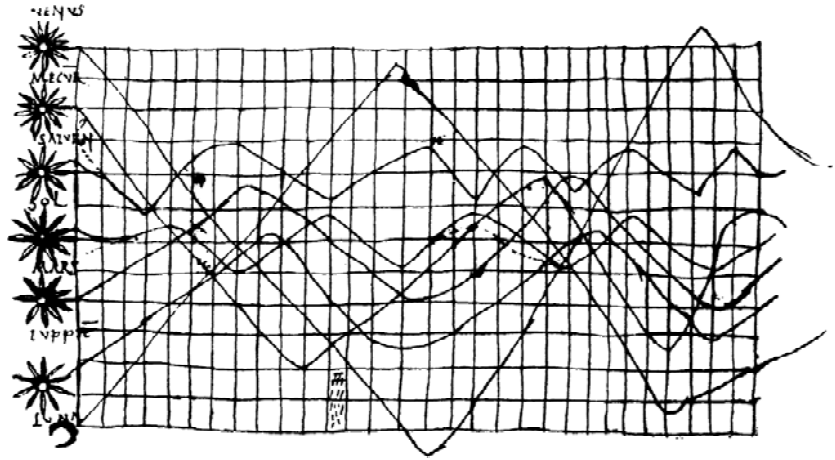

Fig. 1. This is the earliest line graph found in the literature. It divides the 2D plane onto some 30 temporal zones across the *x*-axis and uses horizontal lines to indicate zodiac zones across the *y*-axis. Seven time series were displayed in this chart. Source: [Funkhouser 1936].

Fig. 1 shows a line graph created by an unknown astronomer in the 10th (or possibly 11th) century, depicting the "inclinations of the planetary obits as a function of the time" [Funkhouser 1936]. More line graphs were found in the 17th century records, noticeably the plot of "life expectancy vs. age" by Christiaan Huygens in 1669, and the plot of "barometric pressure vs. altitude" by Edmund Halley in 1686 [Friendly2007]. The 18th and 19th centuries saw the establishment of statistical graphics as a collection of charting methods, attributed to William Playfair, Francis Galton, Karl Pearson and others [Cleveland 1985]. The invention of coordinate papers in the 18th century also helped make line graph a ubiquitous technique in science and engineering.

Today, digitally stored data that captures or exhibits a functional relationship $y = f(x)$ is everywhere. For example, there are thousands or millions of real time data feeds of financial information. Weather stations and seismic monitors around the world generate an overwhelming amount of data in the form of $y = f(t)$. In some cases, we still use line graphs for visualization, and in other cases, we do not. What has been the *most fundamental factor* that makes visualization users choose one visual representation from another? Is it a more quantifiable factor, such as the *number of data series*, the *number of data points per data series*, or another data-centric attribute? Is it a less quantifiable factor such as the *amount or type of insight*, the *amount of cognitive load required*, the *level of aesthetic attraction*, the *type of judgment to be made*, or any other human-centric attribute?

Let us consider why a seismologist uses a seismograph, which is a type of line graph that depicts the measured vibrations over time. (For the convenience of referring, we use female pronouns for the seismologist.) The main task supported by a seismograph is for a seismologist to *make observation*. Her first priority is simply to see, or to know, the data stream in front of her, so she can confidentially say "I have seen the data". She may wish to observe some signature patterns of a potential earthquake, relationships among several data series measured at different locations, anomalies that may indicate malfunction of a device, and so on. The seismologist also uses seismographs as a mechanism of *external memorization*, since they "remember" the data for her. In real time monitoring, she does not have to stare at the seismometer constantly and can have a break from time to time. In offline analysis, she does not need to remember all historical patterns, and can recall her memory by inspecting the relevant seismographs. Viewing seismographs *simulates various thoughts*, such as hypotheses. After observing a certain signature pattern in a seismograph, she may hypothesise that the vibrations would become stronger in the next few hours. While the seismograph advances with newly arrived data, she *evaluates her hypothesis* intuitively. When discussing with her colleagues, she draws their attention to the visual patterns on the seismograph, and explains her hypothesis and conclusion. In other words, she uses the seismograph to *aid her communication* with others.

Perhaps the seismologist does not have to use seismographs. The vibration measures could simply be displayed as a *stream of numbers*; after all viewing these numbers would be more accurate than viewing the wiggly line on a seismograph. Alternatively, to make more cost-effective use of the visual media, the stream of number could be *animated* in real time as a dot moving up and down, accompanied by a precise numerical reading updated dynamically. Let us have a close look at the advantages of a seismograph over a stream of numbers or an animation.

### 2.1 Making Observation

It is not difficult for most people to conclude that viewing a stream of numbers is much slower than viewing a line graph such as a seismograph. Numerous studies in psychology have confirmed this (e.g., [Styles 2006]). The latter often facilitates pre-attentive processing, allowing information to be obtained from a visual medium or environment unconsciously. One might suggest that line graphs make better use of space than a stream data. This is certainly true, but space optimisation cannot be the fundamental factor, as line graphs are not the optimal in terms of space usage in static visualization [Chen and Jänicke 2010]. Furthermore, the *animation* of dots and numbers would offer much better space utilisation.

The difficulty of using *animation* to support tasks of making observations is due to its excessive demand for various cognitive capabilities, including attention and memory. While watching such an animation, it is difficult for a viewer to pay attention to a specific temporal pattern, and almost impossible to have a photographic memory to record a specific set of numbers. Of course, one could view the same animation repeatedly, and would eventually work out interesting patterns in the movement of the dot and the variations of the numbers. It is no doubt much slower than viewing a line graph.

### 2.2 Facilitating External Memorisation

A stream of numbers and an animation can facilitate external memorisation. In fact, almost all digitally-stored data can do so. Hence the question should focus on how fast a form of visual display can facilitate memory recall. Similar to what discussed in Section 2.1, viewing a line graph is much quicker than viewing a stream of numbers, or an animation. For example, if a seismologist tries to recollect her memory about some events taking place over the past few hours, it only takes a few seconds for her to trace her eyes along the seismograph to be reminded about what happened before. It would perhaps take hours to read through thousands of numbers, or to watch the animation repeatedly.

### 2.3 Stimulating Hypotheses and Other Thoughts

While some visualization tasks may involve routine, and perhaps sometimes mundane, observations, others can be highly analytical, involving various aspects of thought process, such as data comprehension, facts deliberation, experience reflection, hypothesis generation, hypothesis evaluation, opinion formulation, and decision making. Many aspects of human thinking are stimulated by visual signals. If we compare a seismograph with a stream of numbers or an animation of dots and numbers, we are interested in which type of visual signal can stimulate more thought, or stimulate a specific aspect of thought faster. To our knowledge, there is yet any reported study on such questions. However, with the availability of technologies, such as functional magnetic resonance imaging (fMRI), we hope that there will be more conclusive answers in the near future.

Nevertheless, there have been many anecdote evidences suggesting that when visualization is appropriately designed to convey overviews and is supported by interaction for details-on-demand exploration, it can simulate hypotheses more effectively.

### 2.4 Evaluating Hypothesis

There is no doubt that hypothesis testing is a critical process in scientific investigation. Whenever applicable and feasible, one should always utilise scientific methods, such as statistical hypothesis testing and Bayesian hypothesis testing.

However, such scientific methods often require a non-trivial amount of time and effort for collecting, processing and analysing data. In practice, visualization is often used as an intuitive form of hypothesis evaluation. For example, in scientific computation, to evaluate a simulation model (which is essentially a hypothesis), scientists visualize the results of simulation and visually compare the results with some ground truth data. Such intuitive evaluation is based on the principle counterfactual reasoning. It is not in any way unscientific. It saves time. In the visualization literature, there have been many case studies that confirm the use of visualization as a tool for hypothesis evaluation.

### 2.5 Disseminating Knowledge

Visualization is used extensively for disseminating knowledge. In fact, this is often mistaken as the main or only function of visualization. In such situations, visualization is a tool for assisting a scientist or scholar in delivering a collection of messages to an audience. These messages may consists of data being visualized, background information to be appreciated, concepts to be comprehended, and opinions to be accepted. Clearly, the person presenting the visualization would like to direct the audience to receive the intended messages as fully, and as fast, as possible.

There is a subtle difference between this visualization task and those in Sections 2.1-2.4. Assessing how well a task is performed is generally difficult in many practical situations. For example, consider a task of making seismological observation. If a visual pattern of a potential risk was not noticed in a seismograph during routine monitoring, unless the risk is actualised, it always seems debatable as to such pattern should be noticed or not. The same paradox can be suggested for external memorisation, hypothesis simulation and hypothesis evaluation. In knowledge dissemination, however, as the person presenting the visualization usually has a set of defined criteria for measuring task performance, he/she can assess the audience to determine whether the intended messages were received. Meanwhile, the time is more a constraint rather than a quality metric, since, for instance, a presentation, a meeting or a lecture is usually time-limited. In such a situation, visualization is often embellished in order to "energise" the messages intended by the presenter.

## 3. SAVING TIME IN DIFFERENT MODES OF VISUALIZATION

The story of line graph in Section 2 highlights the importance of *saving time* in performing a user's tasks. Of course, this is not a new discovery. Amid many "insight-based" definitions, some appreciated the purpose of saving time:

- "Today's researchers must consume ever higher volumes of numbers ... If researchers try to read the data, ... they will take in the information at snail's pace. If the information is rendered graphically, however, they can assimilate it at a much *faster* rate." [Friedhoff and Kiely 1990]
- "One of the greatest benefits of data visualization is the sheer quantity of information that can be *rapidly* interpreted if it is presented well." [Ware 2004]
- "Visual representations and interaction technologies provide the mechanism for allowing the user to see and understand large volumes of information *at once*." [Thomas and Cook 2005]
- "Information visualization promises to help us *speed* our understanding and action in a world of increasing information volumes." [Card, 2007]

### 3.1 What is Visualization Really for?

*DEFINITION.* Visualization (or more precisely, computer-supported data visualization) is a study of transformation from data to visual representations in order to facilitate effective and efficient cognitive processes in performing tasks involving data. The fundamental measure for effectiveness is correctness and that for efficiency is the time required for accomplishing a task.

Note that we choose the verb "accomplish" to emphasise that the task has to be performed to a certain degree of satisfaction before the measure of efficiency becomes meaningful. When the correctness has reached a satisfactory level, or becomes paradoxically difficult to assess (as discussed in Section 2.5), the time required to perform a visualization task becomes the most fundamental factor. Such time is a function of three groups of variables:

(a) *data centric attributes*, such as the size of a dataset, the number of multivariate dimensions, the entropy of the data space, etc.

(b) *human-centric attributes*, such as amount or type of insight to be gained, the type of judgment to be made, the amount of cognitive load required, the level of aesthetic attraction, etc.

(c) *information delivery attributes*, such as the type of medium, the properties of the display device, the type of visual representations, the type of exploration, etc.

In most real world applications, there is usually little flexibility with (a) and (b). Hence choosing the appropriate *information delivery attributes* can be critical to accomplish a task efficiently.

### 3.2 Modes of Visualization

Visualization serves as a medium and a tool for human-human interaction. Let us refer to those who create visualization as **visualization producer** and those who view visualization in order to gain an insight as **visualization consumer**.

In some cases, the producer differs from the consumer. For example, a business analyst, who has a good understanding of a financial data set, creates a collection of charts for a company board meeting; or a teacher, who has a good understanding of a concept, creates an illustration to disseminate his or her knowledge to students. In many cases, the producer is also the consumer. For example, in a visual data mining process, an analyst, who has difficulty to comprehend a complex data set by simply reading the textual or numerical data, interactively explores various visual representations of the data, in order to gain an overview or make a discovery.

Let us consider three types of visualization users: **analyst** $A$, who is a producer as well as a consumer, **presenter** $P$, who is a producer but not a consumer, and **viewer** $V$, who is a consumer but not a producer. Different combinations of analysts, presenter and viewers in visualization processes will usually lead to different styles of human-human interaction. Table 1 lists several typical operational modes of visualization processes.

Table 1: Examples of common modes of human participation in visualization processes.

| Mode | Participants | Example Scenarios |
|---|---|---|
| (1) | $A$ | an analyst works alone. |
| (2) | $A_1, A_2, ..., A_k$ | a team of analysts conduct collaborative visual data mining. |
| (3) | $A, V$ | a personal visualization assistant and a boss. |
| (4) | $P, V$ | a personal tutor and a student. |
| (5) | $P, V_1, V_2, ..., V_n$ | a presenter (or a teacher) and an audience (or students). |
| (6) | $A_1, A_2, ..., A_k,$ $V_1, V_2, ..., V_n$ | a team of analysts carry out visual data mining in real time, while a panel of onlookers eagerly observe the process. |
| (7) | $A, P, V_1, V_2, ..., V_n$ | an analyst working for a domain expert who needs to disseminate his/her research to others |

Most analytical tasks (Sections 2.1-2.4) are likely to be conducted in modes (1), (2), and (3). Only the tasks of knowledge dissemination are normally conducted in modes (4) and (5). Mode (6) is relatively rare, but one can easily imagine that some visualization tasks during disaster management may be performed in this mode. On the other hand, mode (7) is rather common, but often has conflicting requirements between the knowledge dissemination task and those analytical tasks.

### 3.3 Reasoning about Visual Embellishment

Let us revisit the debate about visual embellishment discussed in Section 1. Borgo et al. [2012] reported a study on visual embellishment, which used more conservative stimuli than [Bateman 2010]. It shows that visual embellishment may help information retention in terms of both accuracy of and time required for memory recall. However, this is at the expenses of an increase in the time required for visual search, which is an activity typically taking place in tasks of making observation, hypothesis generation and evaluation. Their study also indicates that visual embellishment may help viewers to grasp concepts that a presenter would like to disseminate, especially when such concepts are embedded among complex information.

Borgo et al [2012] pointed out in their explanation that the finding about the negative impact on visual search tasks provides scientific evidence to indicate some disadvantages of using visual embellishments. In other words, visual embellishment is unlikely to save time for an analyst in performing tasks such as making observation, hypothesis generation and evaluation. They also pointed out that the positive impact on memory and concept grasping should not be generalized to situations where visualizations are created by data analysts for their own use. In other words, the positive impact is relevant mainly to the above-mentioned modes (4) and (5). In addition, they made a connection between their findings and the information theoretic framework of visualization [Chen and Jänicke 2010].

Consider those visualization tasks discussed in Section 2. There are analytical tasks (Sections 2.1-2.4), and dissemination tasks (Section 2.5). Table 2 summarises the main characteristics of these two groups of visualization tasks. Since the majority of work in visualization concerns about analytical tasks, the notion of "saving time" must not sit on the backbench in the definition of visualization. As it implicitly implies the completion of task, it encapsulates the notion of "gaining insight" to a large degree, but not vice versa. Hence, "saving time" is more fundamental.

Table 2: Characteristics of analytical tasks and dissemination tasks in visualization.

|  | **Analytical Tasks** | **Dissemination Tasks** |
|---|---|---|
| Modes of visualization | producer (may) = consumer | producer ≠ consumer |
| Saving time | producer & consumer's time | producer & consumer's time |
| Gaining insight | for producer to gain | for consumer to gain |
| Assessing correctness | relatively difficult | more feasible |
| Using embellishment | usually not helpful | can be helpful |
| Information theory | source encoding | channel encoding |

### 4. HOW "SAVING TIME" MAKES A DIFFERENCE?

One might wonder whether bring the "saving time" emphasis to the frontbench in the definition of visualization has a different implication from those existing definitions given in Section 1. There are indeed some fundamental differences.

#### 4.1 Measurement

Firstly, time is much easier to measure and quantify than insight, knowledge or cognitive load, especially in the case of analytical tasks. In many ways, time may also be easier to measure than information, that is, the quantitative measures used in information theory. While the measurement about insight or cognitive load may be undertaken in a laboratory condition, it is usually far too intrusive for a practical environment. Such a measurement would be uncertain as the measurement introduces a significant amount of artefacts and distortion to a normal cognitive process of gaining insight.

### 4.2 Empirical Studies

Most empirical studies involved measurement of accuracy and response time. It is comforting to know such measurements are not only meaningful, but also fundamental. While we encourage and experiment with other studying methods, it is important not to underestimate the measurement of time.

It is necessary to recognise the limitation of empirical studies in assessing "insight gained", especially when domain-specific knowledge is required. "Insight gained" depends on data as well as existing knowledge of participants. When such knowledge varies dramatically from one person to another, the study results have to be treated with care. Hence empirical studies should focused on fundamental questions in visualization, and have to minimise variables, especially those hard-to-observe and hard-to-control variables such as a priori knowledge and insight to be gained.

### 4.3 Design Optimization

Measuring the time taken to perform a task can often be done seamlessly by a system, subject to the necessary ethical consideration and user consensus. This provides a metric for guiding the optimisation of the design of visual representations, interaction methods, or visual analytics processes. In comparison with other metrics, the time required to perform a task is undoubtedly the most important. It is easier to measure, more objective, and more generic to all types of data, visual designs, systems, tasks and users.

### 4.4 Theory of Visualization

The visualization community has not yet found a theory of visualization that most would agree to be fundamental. The good news is that many researchers are inspired to find such a theory, and some frameworks have been proposed. Any theory of visualization should try to account for the impact of time required for performing or accomplishing visualization tasks.

### 4.5 A Practical Wisdom

Most visualization researchers have had some experience of engaging with scientists or scholars in different disciplines, or potential users from industrial or governmental organizations. Many of us had encountered difficulties in persuading potential collaborators about the merits of using visualization, or the need for developing advanced visual designs and visualization systems. After demonstrating some visualization techniques, typically conversations between a visualization researcher and a potential user might flow like that:

**Potential user** (engagingly): *These pictures are very pretty. We are interested in having such techniques. I wonder how I can justify the costs for developing the system that you proposed.*

**Visualization researcher** (enthusiastically): *As you can see from the demo, visualization enables you to gain new insights from the data, this very much outweighs the development costs.*

The first draft of this article was completed on 20 February 2013

    **Potential user** (doubtfully): *Really, what kind of insights are we talking about?*

    **Visualization researcher** (anxiously): *Patterns.* (Pause, trying to recollect some definitions of visualization.) *Interesting patterns, such as various anomalies, complex associations, warning signs, and potential risks.*

    **Potential user** (hopefully but cautiously): *Can those pictures tell me all these automatically?*

    **Visualization researcher** (truthfully but uneasily): *Not quite automatically. The mapping from data to visual representations will enable you see these patterns more easily and help you to make decisions.*

    **Potential user** (disappointedly): *I can understand my data with no problem. I could not imagine how these pictures can help me make better decisions.*

After a while, some of us learned a wisdom, i.e., never suggesting to potential collaborators that visualization could offer them insight. It is much better to state that visualization could save their time. As Sections 2 and 3 have shown, visualization can indeed save time.

"Gaining insight" has been an elusive purpose of visualization for several decades. It is perhaps the time to invigorate visualization as a scientific discipline by shining the spotlight on a more concrete purpose, that is, *to save the time required for accomplish a visualization task*.